\documentclass{svjour3}                     
\smartqed  
\usepackage{graphicx}
\usepackage{epsfig}
\usepackage{graphicx}
\usepackage{graphicx}
\usepackage{mathptmx}      
\usepackage{latexsym}
 \journalname{Few-Body Systems}

\begin{document}

\title{Van der Waals forces and Photon-less Effective Field Theories}

\author{E. Ruiz Arriola}


\institute{E. Ruiz Arriola \at
           Departamento de F\'isica At\'omica, Molecular y Nuclear,
           Universidad de Granada, E-18071 Granada, Spain.
           \email{earriola@ugr.es}
}

\date{Presented at 21th European Conference On Few-Body Problems In
  Physics: EFB21 \\ 
29 Aug - 3 Sep 2010, Salamanca (Spain)}

\maketitle

\begin{abstract}
In the ultra-cold regime Van der Waals forces between neutral atoms
can be represented by short range effective interactions. We show that
universal low energy scaling features of the underlying vdW long range
force stemming from two photon exchange impose restrictions on an
Effective Field Theory without explicit photons. The role of naively
redundant operators, relevant to the definition of three body forces,
is also analyzed. 
\end{abstract}

\section{Introduction}


From a fundamental QED point of view the underlying mechanism
responsible for van der Waals (vdW) forces corresponds to two
photon-exchange~(see e.g.  \cite{Feinberg:1989ps} and references
therein).  As compared to the short range and exponentially suppressed
chemical bonding forces on sizes about a few Bohr radii, vdW forces
are long range. For interatomic separations $a_B \ll r \ll \hbar c
/\Delta E $, two photons are exchanged in a short time $\sim 2  r/c$
while transitions with excitation energy $\Delta E$ take  
a much larger time $\sim 2 \hbar /\Delta E $, yielding the 
potential 
\begin{eqnarray}
V(r) = - \frac{C_6}{r^6} - \frac{C_8}{r^8} - \frac{C_{10}}{r^{10}}-
\dots 
\label{eq:C6-C8-C10} 
\end{eqnarray}
where $C_n$ are the dispersion coefficients which are accurately known
for many diatomic systems~(see e.g. a compilation in
\cite{Arriola:2009wi}). The vdW {\it length} $R = (M
C_6/\hbar^2)^{\frac14}$ characterizes the size of the forces.
For such potentials, low energy scattering with $ k R \ll 1 $ is
dominated by S-waves which phase-shift, $\delta_0(k)$ , fulfills
the effective range expansion (ERE)~\cite{1963JMP.....4...54L} 
\begin{eqnarray}
k \cot \delta_0 (k)= - \frac1{\alpha_0} + \frac12 r_0 k^2 + v_2 k^4
\log (k^2 R^2) + \dots
\label{eq:ere} 
\end{eqnarray} 
where $\alpha_0$ is the scattering length, and $r_0$ is the effective
range. Note that for this potential the long-range character stars at
${\cal O} (k^4)$ due to the logarithmic piece.

\section{Low energy Scaling of vdW forces}

Remarkably, the effective range was computed
analytically~\cite{Gao:1998zza,PhysRevA.59.1998} when $C_{n \ge 8}=0$
yielding
\begin{eqnarray}
\frac{r_0}{R} & = & \frac{16\,\Gamma\left(5/4\right)^2}{3\pi} 
- \frac{4}{3} \frac{R}{\alpha_0} 
+ \frac{4\Gamma\left(3/4\right)^2}{3\pi} \frac{R^2}{\alpha_0^2} =
1.395 -1.333 \frac{R}{\alpha_0}+ 0.6373  \frac{R^2}{\alpha_0^2}
\label{eq:r0-ggf}
\end{eqnarray}
The scaling of the effective range $r_0$ in the vdW length and the
quadratic $1/\alpha_0$ behaviour is just a particular case of a more
general result~\cite{Cordon:2009wh} (see also~\cite{Cordon:2010gy} in
these proceedings). In fact, the dominance of the leading long
distance $C_6$ term tacitly assumed in
Refs.~\cite{Gao:1998zza,PhysRevA.59.1998} was to be expected {\it a
  priori} by suitably re-writing higher order $C_8$, $C_{10}$
contributions in vdW units
\begin{eqnarray}
 R^2 M V(r) = - (R/r)^6 \left[ 1 + g_1 (R/r)^2 + g_2 (R/r)^4 + \dots \right] 
\end{eqnarray}
where $g_1 \sim 10^{-2}$ and $g_2 \sim 10^{-4}$ for many homonuclear
diatomic systems. Thus one expects that {\it even} for $k R \sim 1 $
higher order corrections are negligible despite the strong divergence
at short distance.  These expectations are indeed met {\it a
  posteriori} on the light of about a hundred calculations based on
phenomenological potentials~\cite{Cordon:2009wh}. This result not only
favours the view that these rather simple approaches based on the
leading vdW forces are phenomenologically sound but also shows that a
huge reduction of parameters takes place suggesting that atoms in the
ultra-cold regime can indeed be handled without much explicit reference
to the underlying electronic structure of atoms. The scaling universal
relation, Eq.~(\ref{eq:r0-ggf}), allows for a quite general discussion
on effective interactions in vdW units, as we advance here.

\section{Effective Short Distance Potentials}

In the ultra-cold regime, i.e. for extremely long de Broglie
wavelengths much larger than the vdW scale, $ \lambda = 1/k \gg R$ ,
one expects the long range character to become largely irrelevant,
keeping the first two terms in Eq.~(\ref{eq:ere}). Thus, one might
want to represent the vdW potential by an effective potential with a
finite range, $r_c$, featuring the truncated ERE, Eq.~(\ref{eq:ere}),
and dismissing any explicit reference to the underlying photon
exchange. However, even at very low energies, causality arguments
provide the shortest possible value for $r_c$, which for vdW forces
yields $r_c> 0.6 R $~\cite{Arriola:2009wi}. Using for illustration a
square well (SW) potential with range $r_c$ and depth $V_0$, $V_{\rm
  eff}(r) = -V_0 \theta (r_c-r)$, one obtains
\begin{eqnarray}
\alpha_0 = r_c - \frac{\tan \sqrt{M V_0} r_c}{\sqrt{M V_0}} \, ,
\qquad r_0 = r_c \left[ 1- \frac{1}{\alpha_0 r_c M V_0} -
  \frac{r_c^2}{3\alpha_0^2} \right] \, .
\end{eqnarray}
Reproducing Eq.~(\ref{eq:r0-ggf}) is not possible for a {\it common}
potential. Indeed, the sign of the $1/\alpha_0^2$ term is just
opposite, so that for small $\alpha_0$ we cannot represent the
interaction by this short range potential. On the contrary, for large
scattering lengths $\alpha_0 \gg R$ we obtain $r_c = 1.395 R $ and
$V_0=\pi^2/(4 r_c^2 M)$.  In terms of volume integrals,
\begin{eqnarray} 
C_0 = \int d^3 x \, V_{\rm eff}(\vec x) \, , \qquad 
C_2 = - \frac16  \int d^3 x \, r^2 \, V_{\rm eff}(\vec x) 
\end{eqnarray}
one gets $ M C_0^{\rm SW}/R=- 14.41 $ and $M C_2^{\rm SW}/R^3=2.80
$. If we use instead a delta-shell (DS) potential $V_{\rm eff}(r) = -
V_0 r_c \delta(r-r_c)$ we get for $\alpha_0 \gg R$ the results $ M
C_0^{\rm DS}/R=- 13.15 $ and $M C_2^{\rm DS}/R^3=2.40$, not far from
the SW estimate. This suggests using a formulation based {\it
  directly} on the constants $C_0$ and $C_2$.  Note that while a $C_4$
exists for these short distance potentials, the original vdW potential
yields a divergence, in harmony with the observation that the ERE for
short range potentials differs at ${\cal O}(k^4)$ from the vdW
expression, Eq.~(\ref{eq:r0-ggf}).

\section{Effective Field Theory}

\begin{figure}
\includegraphics[width=0.46\textwidth]{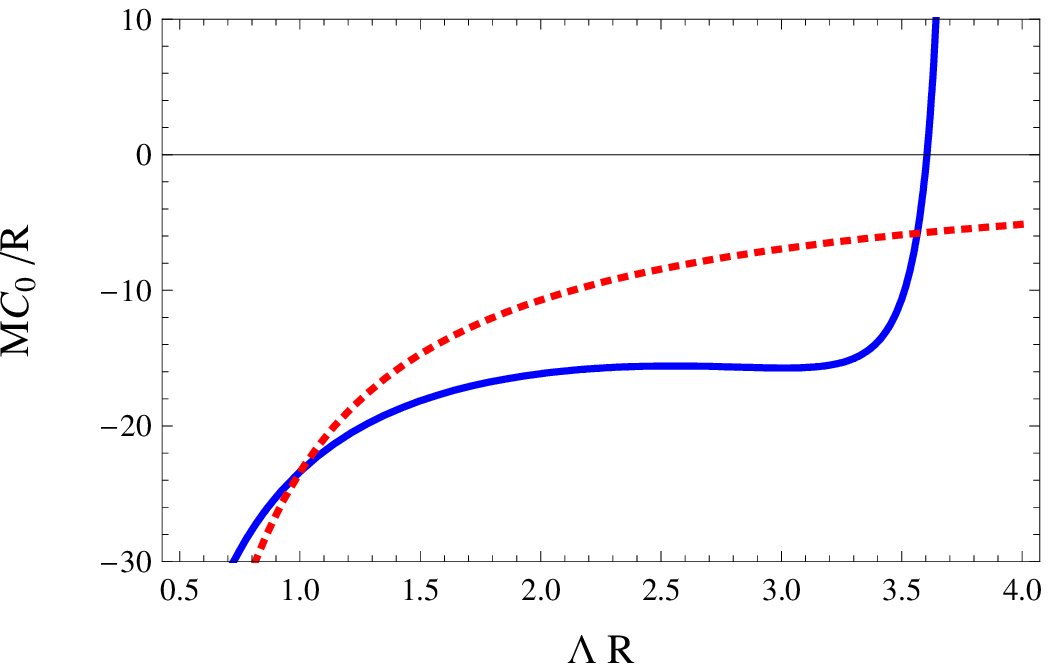} \hskip.5cm
\includegraphics[width=0.46\textwidth]{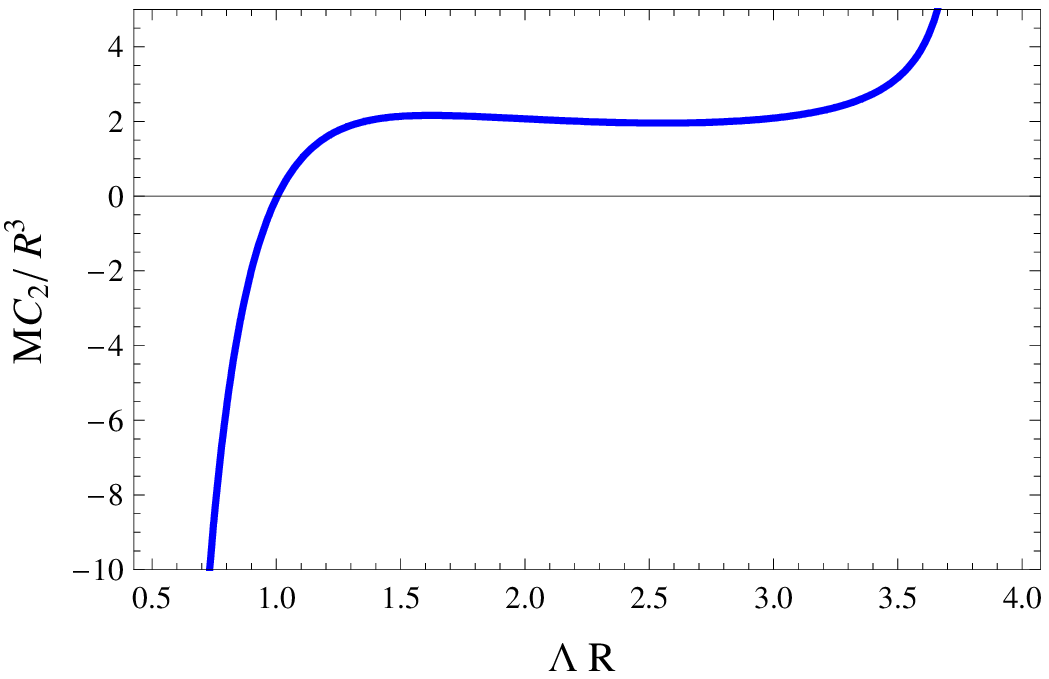}
\caption{Cut-off dependence of the EFT coefficients $M C_0 /R $ when
  $C_2=0$ (dashed red) and $M C_0 /R^3$ and $M C_2 /R^3$ after
  Eqs.~(\ref{eq:r0-EFT}) (full blue) for the case $\alpha_0/R=10$. 
  $R$ is the vdW scale defined as $R = (M C_6/\hbar^2)^{\frac14} $.}
\label{fig:1}       
\end{figure}

The EFT approach has often been invoked to highlight universal
features of ultra-cold few atoms systems (for reviews see
e.g.~\cite{Braaten:2004rn,Platter:2009gz}). We re-analyze it on the
light of the universal and extremely successful scaling relation,
Eq.~(\ref{eq:r0-ggf}).  For definiteness, we consider the Galilean
invariant Lagrangian density~\cite{Braaten:2000eh} expanded in
composite Bosonic spinless field operators with increasing energy
dimensions and including multi-particle interactions,
\begin{eqnarray}
{\cal L} = \psi^\dagger \left( i \partial_t + \frac{\nabla^2}{2m}
\right) \psi - \frac{C_0}2 (\psi^\dagger \psi )^2 - \frac{C_2}2 [
  \nabla (\psi^\dagger \psi )]^2 -\frac{D_0}6 (\psi^\dagger \psi )^3 +
\dots
\end{eqnarray}
Here $C_0$, $C_2$ and $D_0$ are low energy constants which are fixed
from few body dynamics. Using Feynman rules in the two-body sector one
derives a scale dependent and momentum truncated self-adjoint
pseudo-potential in the CM system ($\vec k$ and $\vec k'$ are relative
momenta)
\begin{eqnarray}
\langle \vec k' | V| \vec  k \rangle 
= \left[ C_0 + C_2 (\vec k^2 +\vec k'^2) + \dots \right]
\theta (\Lambda -k) \theta (\Lambda -k') \, . 
\label{eq:pot-mom}
\end{eqnarray}
The cut-off $\Lambda$ is introduced here to handle the power divergent
integrals arising in the scattering problem, which in terms of the
Lippmann-Schwinger (LS) equation becomes
\begin{eqnarray}
\langle \vec k' | T | \vec k \rangle = \langle \vec k' | V| \vec k
\rangle +  M \int\frac{d^3 q }{(2\pi)^3} \frac{ \langle \vec
  k' | V| \vec q \rangle \langle \vec q | T| \vec k \rangle
}{p^2-q^2+i 0^+} \, , 
\label{eq:LS-mom} 
\end{eqnarray}
implementing unitarity for $p \le \Lambda$. Using the potential of
Eq.~(\ref{eq:pot-mom}) the LS Eq.~(\ref{eq:LS-mom}) reduces to a
system of algebraic equations which solution is well known (see e.g.
Ref.~\cite{Entem:2007jg}) yielding
\begin{eqnarray} 
-\frac1{\alpha_0 \Lambda } &=& \frac{4 \left(-2 c_2^2+90 \pi^4
+15 (3 c_0+2 c_2) \pi ^2\right)}{9 \pi
  \left(c_2^2 -10 c_0 \pi ^2 \right)} \, , 
\label{eq:r0-EFT}
\\ r_0 \Lambda &=& \frac{16 \left(c_2^2+12 \pi ^2 c_2+9
  \pi ^4\right)}{\pi \left(c_2+6 \pi ^2\right)^2} -\frac{12
  c_2 \left(c_2+12 \pi ^2\right)}{\left(c_2+6 \pi
  ^2\right)^2} \frac1{\alpha_0 \Lambda} +\frac{3 c_2 \pi
  \left(c_2+12 \pi ^2\right)}{\left(c_2+6 \pi
  ^2\right)^2} \frac1{\alpha_0^2 \Lambda^2} \, , \nonumber 
\end{eqnarray} 
where $c_0 = M \Lambda C_0$, $c_2 = M \Lambda^3 C_2$.  By eliminating
$C_0$ in terms of $\alpha_0$ we have written $r_0$ in a form similar
to Eq.~(\ref{eq:r0-ggf}).  This leads for any cut-off $\Lambda$ to the
mapping $(\alpha_0,r_0) \to (C_0, C_2) $.  For $C_2=0$ one gets $r_0=
4/\pi \Lambda$ which for $\alpha_0 \gg R $ yields $\Lambda R = 0.91$
and $M C_0/R=-21.6$ from matching the scattering length and the
effective ranges $r_0^{\rm vdW}= r_0^{\rm EFT}$.  The cut-off
dependence for $C_2 \neq 0$ can be looked up at Fig.~\ref{fig:1} in
vdW units and for the specific case $\alpha_0/R=10$ where a weakly
bound state takes place.  As we see there is a clear stability plateau
in the region $\Lambda \sim \pi /(2R)$ illustrating the basic point of
the EFT; low energy physics is cut-off independent within a given
cut-off window which does not resolve length scales shorter than the
vdW scale. Numerically we get $M C_0 /R \sim - 15 $ and $M C_2 /R^3
\sim 2 $ for $\Lambda \sim \pi /(2R)$, in agreement with the previous
SW and DS analysis.  The values of $\Lambda$ where the EFT low energy
parameters diverge correspond to an upper bound above which $C_0$ and
$C_2$ become complex, violating the self-adjointness of the potential
~\cite{Arriola:2009wi} and the Lagrangian, ${\cal L} (x)\neq {\cal
  L}^\dagger (x)$.  Thus, off-shell two-body unitarity and hence
three-body unitarity are jeopardized for $\Lambda R \ge 4 $, despite
the phase shift being real and on-shell unitarity being fulfilled.

%



Direct inspection shows that a perfect matching between the vdW and
the EFT effective ranges, Eq.~(\ref{eq:r0-ggf}) and
Eq.~(\ref{eq:r0-EFT}) for {\it any} $\alpha_0$ is not possible.  So we
try out including redundant operators which are usually
discarded~\cite{Braaten:2000eh} but are needed to guarantee off-shell
renormalizability of the LS equation~\cite{Harada:2005tw}.  A Galilean
invariant term of the form $\Delta {\cal L}= - \frac12 C_2'
(\psi^\dagger \psi ) \left[\psi^\dagger \left( i \partial_t +
  \nabla^2/2m \right) \psi \right] $ is formally redundant since it
can be eliminated by a field transformation $\psi \to \psi + \frac14
C_2' \psi (\psi^\dagger \psi )$ which generates additional three body
forces as well. The new term adds a correction $\Delta V = C_2' (2
p^2- \vec k^2-\vec k'^2)/2 $ to the potential, Eq.~(\ref{eq:pot-mom}),
vanishing on-shell.  Solving the LS equation and eliminating $C_0$ in
terms of $\alpha_0$ yields
\begin{eqnarray}
r_0 \Lambda &=& \frac{16 \left((c_2'-2 c_2)^2+36 \pi
  ^4+6 (8 c_2-c_2') \pi ^2\right)}{\pi \left(-2
  c_2+c_2'-12 \pi ^2\right)^2} \nonumber \\ 
&-&\frac{12 \left( (c_2'-2
  c_2)^2+48 c_2 \pi ^2\right)}{ \left(-2
  c_2+c_2'-12 \pi ^2\right)^2}\frac1{ \Lambda \alpha_0}+\frac{3
  \pi \left((c_2'-2 c_2)^2+48 c_2 \pi
  ^2\right)}{\left(-2 c_2+c_2'-12 \pi ^2\right)^2} 
\frac1{  \Lambda^2 \alpha_0^2} \, , 
\end{eqnarray}
where $c_2' = M \Lambda^3 C_2'$ appears through the combination $C_2'-
2C_2$ which {\it cannot} be completely eliminated by making $C_2 \to
C_2+ \frac12 C_2'$.  Note the accidental correlation $-4/\pi$ between
the second and the third coefficients holding regardless on the
particular regularization method. Perfect matching can only be
achieved with complex coefficients. Minimizing the difference between
$r_0^{\rm EFT} $ and $r_0^{\rm VdW}$ provides a reasonable range
$\Lambda R = 1.6-1.8 \sim \pi/2 $.  As we can see, universal two-body
scaling features encoded in Eq.~(\ref{eq:r0-ggf}) and exhibiting the
underlying vdW (two photon exchange) nature of interactions impose
severe restrictions on the EFT solution with no explicit photonic
degrees of freedom and distinguish between naively unitarily
equivalent Hamiltonians mixing different particle number (see
e.g. Ref.~\cite{Furnstahl:2000we}). Therefore, these limitations are
expected to play a role in the EFT analysis of  three-body forces.

\begin{acknowledgements}
I thank A. Calle Cord\'on for collaboration in
\cite{Arriola:2009wi,Cordon:2009wh} .  Work  supported by
  Ministerio de Ciencia y Tecnolog\'\i a under Contract
  no. FIS2008-01143/FIS and Junta de Andaluc{\'\i}a grants no.
  FQM225-05
\end{acknowledgements}

\bibliographystyle{spphys}       


\end{document}